\documentstyle[epsfig]{aipproc}

\newcommand{\ga}{\mathrel{\hbox{\rlap{\lower.55ex \hbox {$\sim$}}
                   \kern-.3em \raise.4ex \hbox{$>$}}}}
\newcommand{\la}{\mathrel{\hbox{\rlap{\lower.55ex \hbox {$\sim$}}
                   \kern-.3em \raise.4ex \hbox{$<$}}}}

\begin{document}
\title{Convection in radiatively inefficient \\ back hole accretion flows}

\author{Igor V. Igumenshchev$^*$ and Marek A. Abramowicz$^{\dagger}$}
\address{$^*$Institute of Astronomy, 48 Pyatnitskaya Ulitsa,
109017 Moscow, Russia\\
$^{\dagger}$Department of Astronomy \& Astrophysics, G\"oteborg University \&
Chalmers University of Technology,
412-96~G\"oteborg, Sweden}

%\lefthead{LEFT head}
%\righthead{RIGHT head}
\maketitle

\begin{abstract}
Recent numerical simulations of radiatively inefficient accretion flows
onto compact objects have shown that convection is a general feature
in such flows. Dissipation of rotational and gravitational energies
in the accretion flows results in inward increase of entropy and
development of efficient convective motions. Convection-dominated
accretion flows (CDAFs) have a structure that is modified significantly
in comparison with the canonical advection-dominated and Bondi-like
accretion flows. The flows are characterized by the flattened radial
density profiles, $\rho(R)\propto R^{-1/2}$, and have reduced mass accretion
rates. Convection transports outward a significant amount of the released
binding energy of the accretion flow. We discuss basic dynamical
and observational properties of CDAFs using numerical models and
self-similar analytical solutions.
\end{abstract}

\section*{INTRODUCTION}

Observations of accreting black holes of different mass,
from the stellar mass black holes to the supermassive black holes in
the center of galaxies, show impressive similarities of data, 
which point at identical physical processes in accreting plasma. 
Also, black hole candidates
demonstrate a great variety of physical conditions in the flows
and, possibly, existence of a variety of accretion regimes.
Existing theories describe
different regimes of black hole accretion flows, 
which can be realized under different physical conditions.
If matter accretes with
a low specific angular momentum, $j\ll R_g c$, 
it forms spherical flows, which are described by Bondi solution 
\cite{bondi52} in the case of adiabatic flows.
Here $R_g$ is the gravitational radius of the black hole, and
$c$ is the speed of light.
If matter has a large specific angular momentum,
$j\gg R_g c$, then accretion disks are formed.
The structure of accretion disks crucially depends on the efficiency of
radiative cooling. If matter radiates efficiently,
i.e. the radiative cooling time
is shorter than the accretion time, $t_{rad}\ll t_{accr}$, the disks
are geometrically thin, $H\ll R$, where $H$ is the scale height of the flow,
and $R$ is the corresponding radius.
The radial structure of such disks is described by the Shakura \& Sunyaev
model \cite{ss73}. In the case of radiatively inefficient flows,
$t_{rad}\gg t_{accr}$, the internal energy of accreting matter is
close to the virial energy, and the formed disks are geometrically thick,
$H\sim R$. The thick disks can be formed in two limiting cases of very high,
$\dot{M}\gg L_{Edd}/c^2$, and very low,
$\dot{M}\ll L_{Edd}/c^2$, accretion rates, where $L_{Edd}$ is the Eddington
luminosity of the black hole. 
In the former case, the optical depth of
accretion flow is very large, and photons, which carry most of
the internal energy, are trapped inside the inflowing
matter and can not be radiated away. In the latter case, the accreting
plasma is very diluted, so that different radiative mechanisms are inefficient
to cool plasma on the accretion time scales. A model which describes
the geometrically thick accretion disks 
was called 
the advection-dominated accretion flow (ADAF) and attracted
a considerable attention during the last two decades [3--10].
%(\cite{ich77}, \cite{gil81}, \cite{bm82},
%\cite{acls88}, \cite{ny94}, \cite{acklr95},
%\cite{ny95a}, \cite{ny95b}).
The main feature of ADAFs is that most of the locally released 
gravitational and rotational
energies of accretion flow is advected inward in the form of
the gas internal energy, and the latter is finally absorbed by the black hole.
The recent interest to ADAFs was generated mostly by
attempts using this model to explain the low-luminosity X-ray objects of
both galactic and intergalactic nature 
(for recent reviews see [11,12]).
%\cite{kfm98} and \cite{abp98}).

\section*{ADVECTION-DOMINATED FLOWS}

Properties of ADAFs could be better understood by analyzing the self-similar 
solution of the height-integrated hydrodynamical equations \cite{ny95b}. 
The solution depends on two parameters, the viscosity parameter $\alpha$
and the adiabatic index $\gamma$, and satisfies the following scalings
for the density $\rho$, the radial velocity $v_R$, the angular velocity
$\Omega$, the isothermal sound speed $c_s$ and $H$:
\[
 \rho(R)\propto R^{-3/2},
\]
\[
 v_R(R)\propto R^{-1/2},
\]
\begin{equation}
 \Omega(R)\propto R^{-3/2},
\end{equation}
\[
 c_s(R)\propto R^{-1/2},
\]
\[
 H(R)=c_s/\Omega_K\propto R,
\]
where $\Omega_K$ is the Keplerian angular velocity.
Because of neglection of the radiative energy losses of the flow,
the flow structure is independent of the mass accretion rate $\dot{M}$.
The accretion rate determines only scales of density through the relation
$\rho=\dot{M}/(4\pi R H v_R)$.

Two inconsistencies were found related to the model of ADAFs.
The first is that ADAFs must be convectively unstable [4,5,7].
%(\cite{gil81}, \cite{bm82}, \cite{ny94}). 
The second 
is connected to the positiveness of ``Bernoulli parameter'' in
ADAFs \cite{ny95a},
\[
 Be={v_R^2\over 2} + W - {GM\over R} > 0,
\]
where $W$ is the specific enthalpy. A hypothesis  was
proposed that the latter problem could be solved by 
assuming the presence of powerful bipolar outflows in the accretion flows
[13,14].
%\cite{xc97}, \cite{bb99}.
However, according to the later investigations [15,16],
the hypothesis met some difficulties.
%\cite{ogil99}, \cite{ali00}.
The consequences of the earlier mentioned inconsistencies for ADAFs 
was understood
after performing two- and three-dimensional hydrodynamic simulations 
of the inefficiently radiative
accretion flows [17--22].
%\cite{ica96}, \cite{ia99}, \cite{spb99}, \cite{igum00}, 
%\cite{ian00}, \cite{ia00}.
The simulations have revealed a new accretion regime of inefficiently radiated
plasma, in which convection plays a dominant role determining structure and
dynamics of the flow. Such a regime we shall refer as CDAF, the
convection-dominated accretion flow.

\section*{CONVECTION INSTABILITY}

Behavior of convective blobs in non-rotating medium
depends sensitively on the superadiabatic gradient
\[
\Delta\nabla c_s^2=-c_s^2{d\over dR} \ln\left({P^{1/\gamma}\over\rho}\right),
\]
which determines the Brunt-V\"ais\"al\"a frequency $N$ given by
\[
N^2=-{g_{eff}\over c_s^2}\Delta\nabla c_s^2,
\]
where $g_{eff}=-(1/\rho)(dP/dR)$ is the radial effective gravity. 
When $N^2$ is positive, 
perturbations of the blobs have an oscillatory 
behavior with the frequency $N$
and the medium is convectively stable. However, when $N^2$
is negative, perturbations have a runaway
growth, leading to convection. Convection is present whenever
$\Delta\nabla c_s^2$ is positive, i.e. when the entropy gradient,
given by
\[
T{ds\over dR}=-{\gamma\over\gamma -1} \Delta\nabla c_s^2,
\]
is negative. This is the well-known Schwarzschild criterion.

When there is rotation, a new frequency enters the problem and convection 
is no longer determined purely by the Brunt-V\"ais\"al\"a frequency.
In this case, the effective frequency $N_{eff}$ of convective blobs
is given by 
\[
N_{eff}^2=N^2 + \kappa^2,
\]
where $\kappa$ is the epicyclic frequency; for $\Omega\propto R^{-3/2}$,
we have $\kappa=\Omega$. Again, when $N_{eff}^2$ is negative,
the rotating flows are convectively unstable.

When convection is developed, the
convective blobs carry outward some amount of thermal energy. Under certain
approximations (see details in \cite{nia00}) the convective energy flux can be
expressed in the form,
\begin{equation}
F_{conv}=-\nu_{conv}\rho T{ds\over dR},
\end{equation}
where $s$ is the specific entropy, $T$ is the temperature,
$\nu_{conv}$ is the diffusion coefficient, defined by
$\nu_{conv}=(L_M^2/4)(-N_{eff}^2)^{1/2}$, and $L_M$ is the characteristic
mixing-length. The coefficient $\nu_{conv}$ can also be expressed in the
$\alpha$-parameterization form, 
$\nu_{conv}=\alpha_{conv} c_s^2/\Omega_K$, 
where $\alpha_{conv}$ is a dimensionless
parameter that describes the strength of convective diffusion;
this parameter is similar to the usual Shakura \& Sunyaev $\alpha$.

In the case of the self-similar ADAF solution (1),
calculations lead to the conclusion that $N_{eff}^2<0$, i.e. ADAFs
are convectively unstable and must have the outward energy fluxes
defined by equation (2).

\section*{Numerical results}

To describe the results of numerical simulations of radiatively
inefficient accretion flows, we shall mainly follow [21,22].
In these works the accretion flows were simulated by solving
the nonrelativistic, time-dependent Navier-Stokes equations
with thermal conduction,
\[
{d\rho\over dt}+\rho\nabla{\bf v}=0,
\]
\begin{equation}
\rho{d{\bf v}\over dt}=-\nabla P+\rho\nabla\Phi+\nabla{\bf \Pi},
\end{equation}
\[
\rho{de\over dt}=-P\nabla{\bf v}-\nabla{\bf q}+Q,
\]
where $e$ is the specific internal energy, $\Phi$ is the gravitational
potential of the black hole,
${\bf \Pi}$ is the viscous stress tensor with all components included,
${\bf q}$ is the heat flux density due to thermal conduction and
$Q$ is the dissipation function. No radiative cooling was assumed and
the ideal gas equation of state, $P=(\gamma-1)\rho e$, was adopted.
Only the shear viscosity was considered, 
\begin{equation}
\nu=\alpha c_s^2/\Omega_K,
\end{equation}
where $\alpha$ is a constant, $0<\alpha\le 1$.
%Details of numerical codes used are described in \cite{ia00}, \cite{ian00}.

In the simulations, it was assumed that mass is steadily injected into the
calculation domain from an equatorial torus near the outer boundary.
Matter is injected with almost Keplerian angular momentum. Owing to viscous
spread, a part of the injected matter moves inward and forms an accretion
flow. The computations were started 
from an initial state in which there is a small
amount of mass in the domain. After a time comparable to the viscous
timescale, the accretion flow achieves a quasi-stationary behavior.

\begin{figure*}[b!] % fig 1
\centerline{\epsfig{file=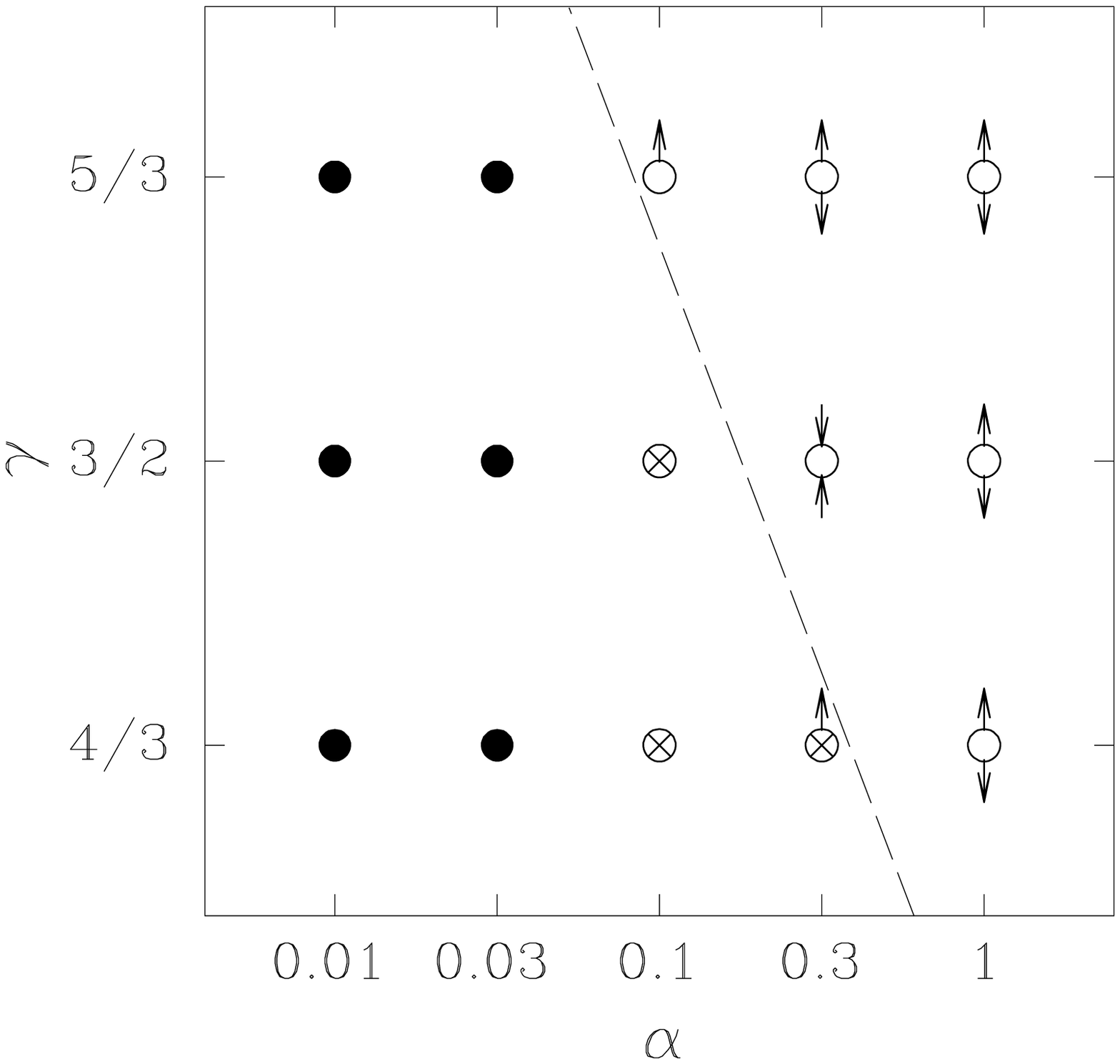,height=3in,width=3in}
\hbox to 0.0in {}\epsfig{file=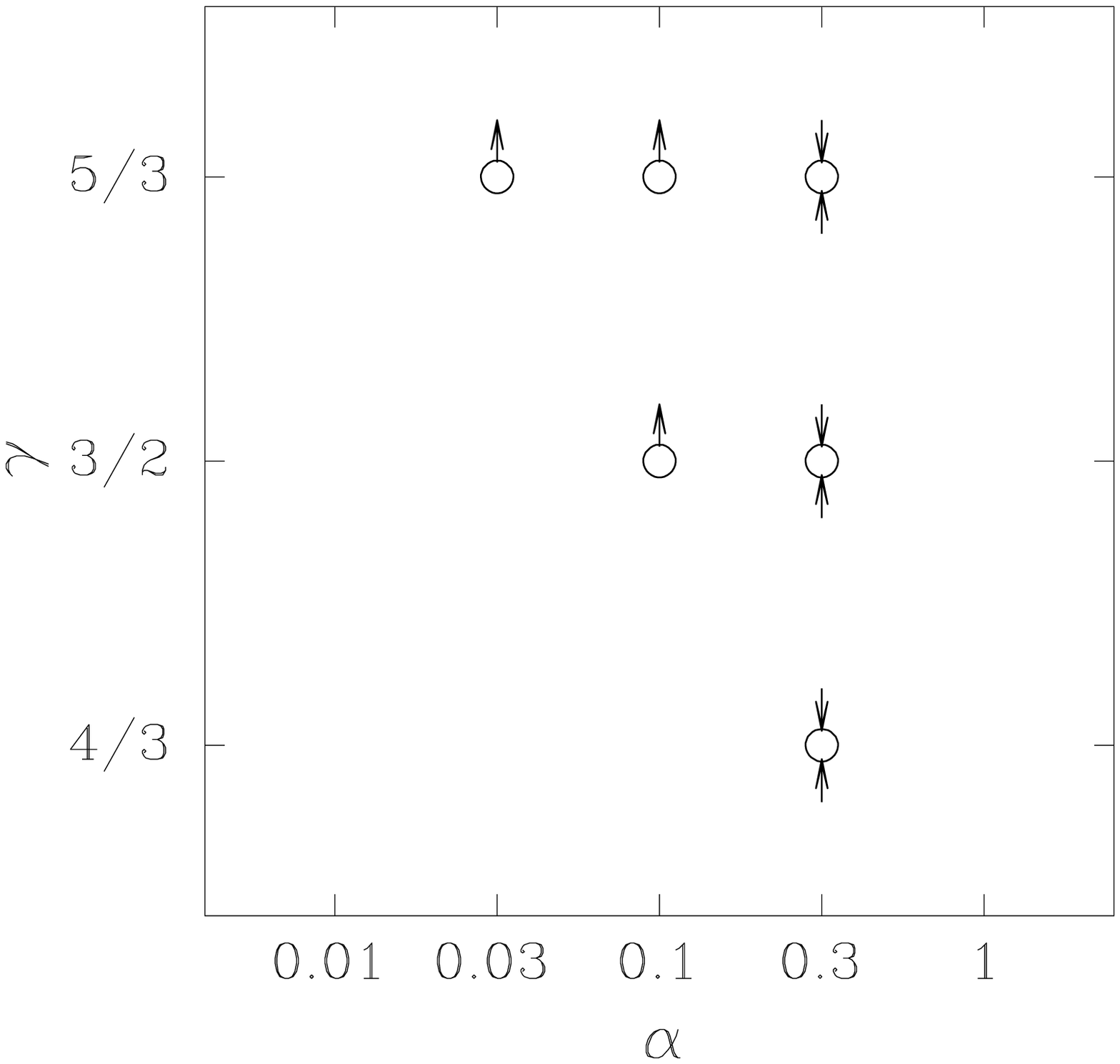,height=3in,width=3in}}
\vspace{10pt}
\caption{Properties of two-dimensional axisymmetric models of accretion flows.
Models with no thermal conduction are shown in the left panel.
Models with thermal conduction (Prandtl number ${\bf Pr}=1$)
are shown in the right panel. Each circle represents a model in the
($\alpha$,$\gamma$) parameter space. The empty circles correspond 
to laminar flows, the crossed circles represent unstable models
with large-scale ($\sim R$) meridional circulations and
solid circles indicate convective models.
Two outward directed arrows correspond to bipolar outflows, whereas
one arrow corresponds to a unipolar outflow. Two inward directed arrows
correspond to a pure inflow. The dashed line on the left-hand panel
approximately separates regions of convectively stable/unstable flows.}
\label{fig1}
\end{figure*}

Results obtained by using two-dimensional axisymmetric
simulations are summarized in Figure~1. Four types of accretion flows
can be distinguished from the models with no thermal conduction
(see Figure~1, left panel).

1. Convective flows. For a very small viscosity, $\alpha\la 0.03$,
accretion flows are convectively unstable. Axially symmetric convection
transports the angular momentum {\it inward} rather than outward.
Convection governs the flow structure,
making a flattened density profile, $\rho(R)\propto R^{-1/2}$, with
respect to the one for ADAFs. Convection transports a significant
amount (up to $\sim 1\%$) of the dissipated binding energy outward.
No powerful outflows are present.

2. Large-scale circulations. For a larger, but still small viscosity,
$\alpha\sim 0.1$, accretion flows could be both stable or unstable
convectively, depending on $\alpha$ and $\gamma$. The flow pattern
consists of the large-scale ($\sim R$) meridional circulations.
No powerful unbound outflows are presented. In some respect this
type of flow is the limiting case of the convective flows in which
the small-scale motions are suppressed by larger viscosity.

3. Pure inflows. With an increasing viscosity, $\alpha\simeq 0.3$,
the convective instability dies off. Some models (with $\gamma\simeq 3/2$)
are characterized by a pure inflow pattern, and agree in many respects
with the self-similar ADAFs. No outflows are present.

4. Bipolar outflows. For a large viscosity, $\alpha\simeq 1$,
accretion flows differ considerably from the simple self-similar models.
Powerful unbound bipolar outflows are present.

\begin{figure}[b!] % fig 2
\centerline{\epsfig{file=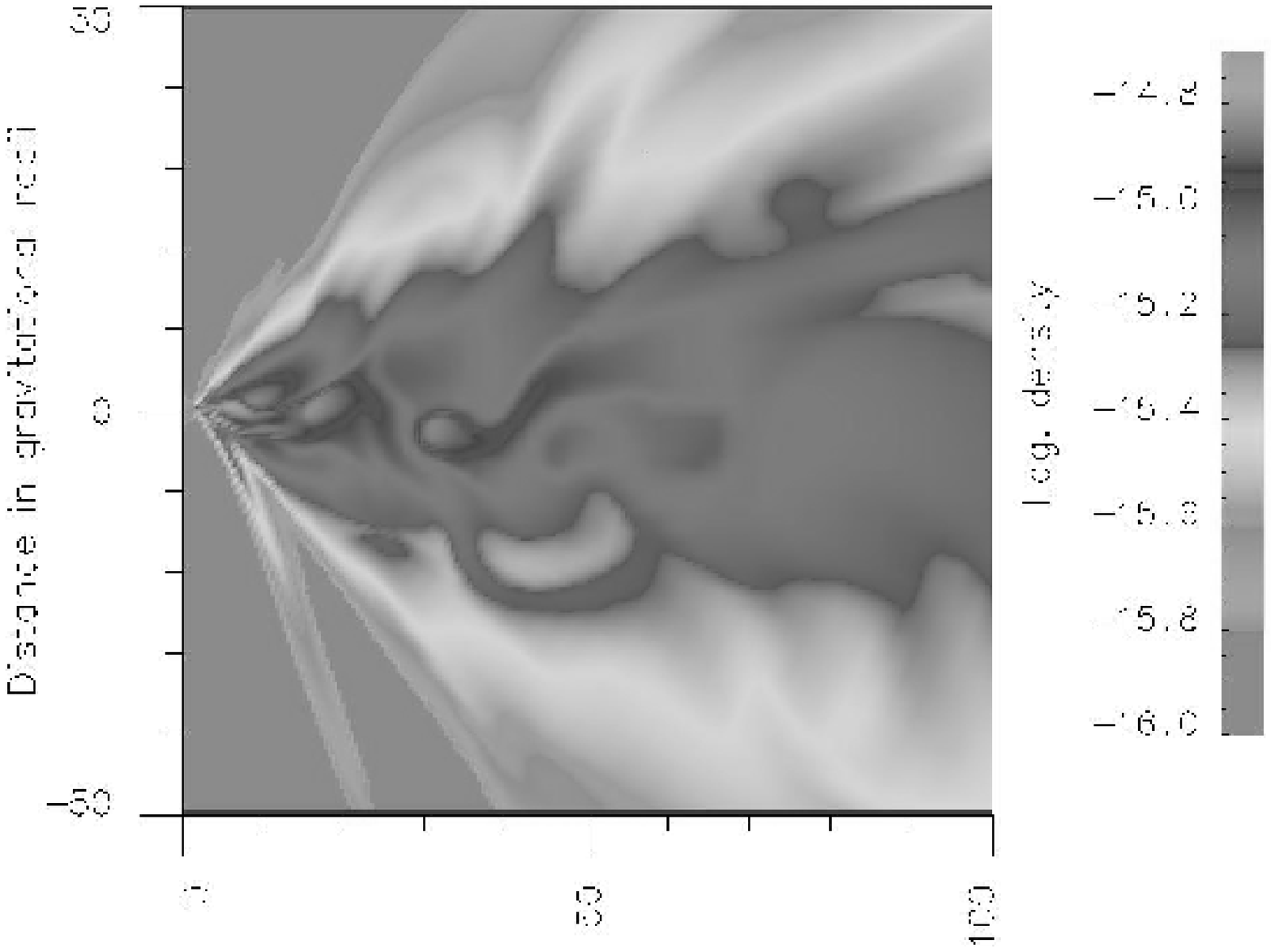,height=3.21in,width=4.3in}}
\vspace{18pt}
\centerline{\epsfig{file=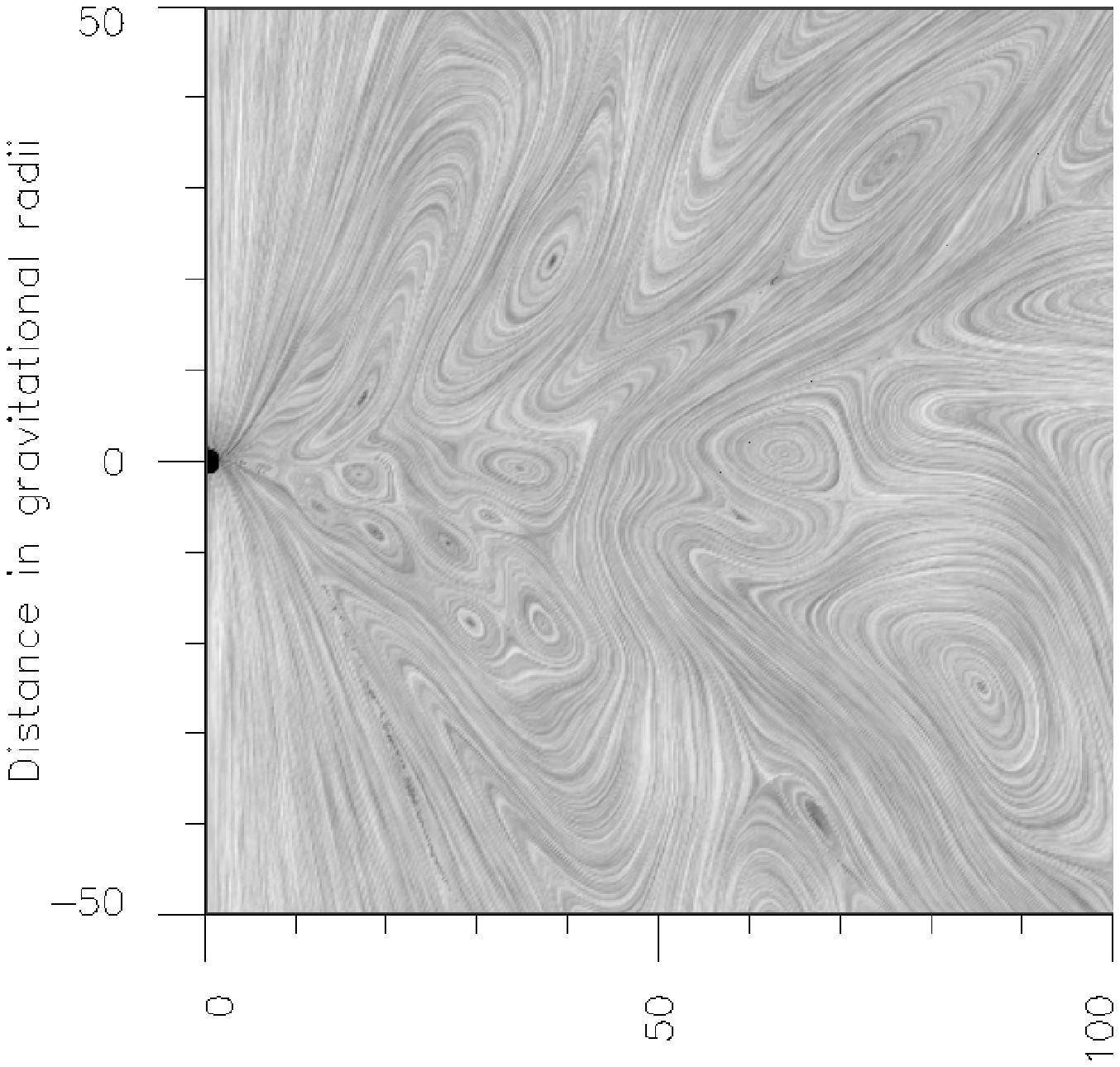,height=3.19in,width=3.375in}\hbox to 0.93in {}}
\vspace{10pt}
\caption{Snapshots of density distribution (upper panel) and
streamlines (lower panel)
from two-dimensional model of CDAF with $\alpha=0.01$ and $\gamma=5/3$,
in the meridional cross-section. The black hole is located at the origin.
The flow pattern is highly time dependent and consists of 
numerous vortices of different spatial scales. The variability of
flow pattern is accompanied by density variations and results in variability
of the accretion rate.}
\label{fig2}
\end{figure}

\begin{figure}[b!] % fig3 
\centerline{\epsfig{file=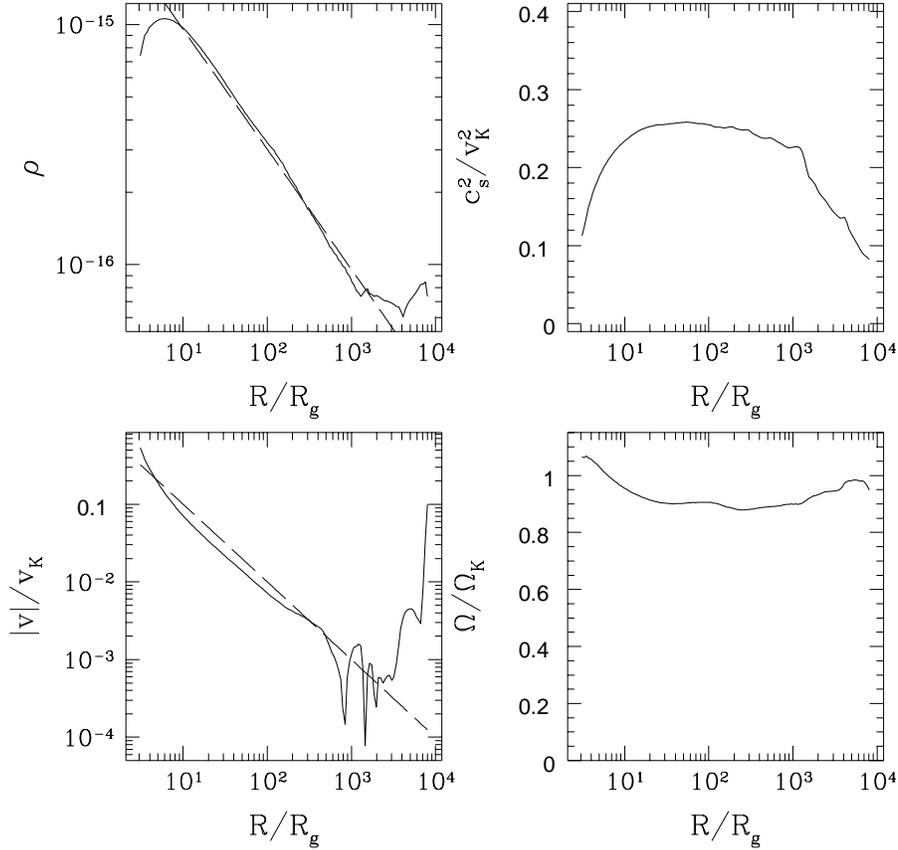,height=5in,width=5in}}
\vspace{10pt}
\caption{Selected properties of two-dimensional 
axisymmetric model with $\alpha=0.01$
and $\gamma=5/3$ [23]. 
All quantities have been averaged over polar angle 
$\theta$ and time. 
Except near the boundaries ($R<10 R_g$ and $R>10^3 R_g$),
the profiles of $\rho$, $c_s^2/v_K^2$, and $\Omega/\Omega_K$ show 
the power-law behaviors predicted by the self-similar
convective envelope solution.
In the upper left panel the dashed line corresponds
to the analytical scaling $\rho\propto R^{-1/2}$.
Similarly, in the lower left panel the dashed line corresponds to the
expected scaling $v\propto R^{-3/2}$ for CDAFs.}
\label{fig3}
\end{figure}

Figure~2 illustrates the density distribution and flow pattern in
the low viscosity models. The models show complicated time-dependent behavior
with numerous vortices and circulations, and with density fluctuations.
However, the time-averaged flow patterns are smooth and do not
demonstrate small-scale features. Figure~3 shows typical behavior
of radial profiles
of variety of time- and angle-averaged variables
in convective models. Except near the inner and outer boundaries 
($R<10 R_g$ and $R>10^3 R_g$), 
the profiles of $\rho$, $c_s$, $v$ and $\Omega$ show
prominent power-law behaviors.

\begin{figure}[b!] % fig4 
\centerline{\epsfig{file=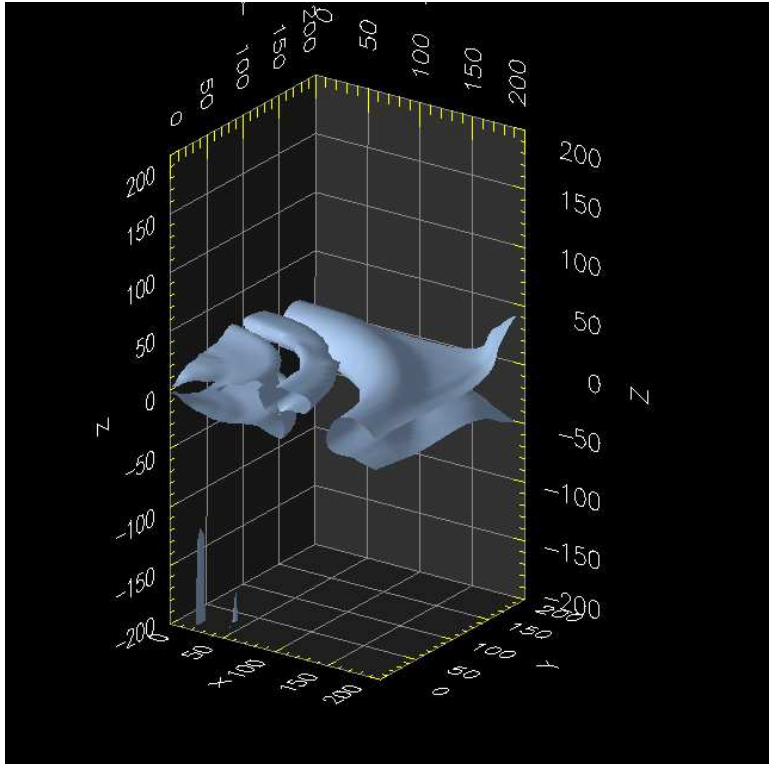,height=4in,width=4in}}
\vspace{10pt}
\caption{Snapshot of surfaces of a constant density
in three-dimensional model of CDAF with $\alpha=0.01$ and $\gamma=5/3$.
Only a quarter of the innermost region of the full domain is shown.
Accretion matter rotates around $z$-axis.
The black hole is located at the origin.
All axes are labeled in the units of $R_g$.
One can clearly see that density perturbations, associated
with motion of convective blobs in the accretion flow, 
have axially symmetric form with respect to the axis of rotation.}
\label{fig4}
\end{figure}

Two-dimensional models have demonstrated that the 
($r$,$\varphi$)-component of the volume averaged Reynolds stress tensor
takes negative values in the case of convective flows.
This means that axisymmetric
convection turbulence transports the angular momentum inward.
Indeed, if there are no azimuthal gradients of pressure, 
the turbulence attempts
to erase the angular momentum gradient, which for the Keplerian-like
angular velocity, $\Omega\propto R^{-3/2}$, means that the 
angular momentum transports inward. 
However, this is a property of axisymmetric flows.
Does convection in real three-dimensional
flows move angular momentum inward? Do such flows have significant
azimuthal perturbations which destroy axisymmetry? 
The answer to this question has been found with the help of
fully three-dimensional simulations,
without the limiting assumption of axisymmetry \cite{ian00}. 
The simulations have demonstrated
a good qualitative and quantitative coincidence of the results 
of two- and three-dimension modeling. It was demonstrated that in the 
differentially rotating flows 
with nearly Keplerian angular velocities,
all azimuthal gradients are efficiently washed out, leading to
almost axisymmetrical structure of three-dimensional flows. 
Figure~4 illustrates
axisymmetric structure of the low-viscosity accretion flow 
obtained in three-dimensional simulations.

All low-viscosity models have negative volume-averaged $Be$ in all ranges 
of the radii. Only temporal convective blobs and narrow regions at the 
disk surfaces show positive $Be$. It seems that the outward directed
convection energy transport
provides an additional cooling mechanism
of the flows that always results in averaged $Be<0$, 
opposite to positive $Be$ obtained in ADAFs.

In several axisymmetric simulations, effects 
of turbulent thermal conduction have been studied.
It was found that the conduction has an important influence
on the flow structure, but it does not introduce a new type of flow
in addition to those already discussed (see Figure~1, right panel). 
The conduction leads to
a suppression of small-scale convection in the low-viscosity models. 
In the moderate-viscosity models the thermal conduction acts as a cooling
agent in the outflows, reducing or even suppressing them.

Obtained properties of CDAF can have important implications for the
spectra and luminosities of accreting black holes [20,22,23]. Indeed, since
$\rho\propto R^{-1/2}$ and $T\propto R^{-1}$, the bremsstrahlung cooling
rate per unit volume varies as 
$Q_{br}\propto \rho^2 T^{-1/2}\propto R^{-3/2}$.
After integration of $Q_{br}$ over bulk of the flow one obtains the
bremsstrahlung luminosity $L_{br}\propto R$.
This means that CDAFs mostly radiate on the outside, whereas
most of the radiative energy of ADAFs  comes from the innermost
region \cite{ny95b}.

\section*{SELF-SIMILAR CONVECTIVE ENVELOPE}

Numerical simulations of convective accretion flows 
have been understood qualitatively after construction of
analytical self-similar solutions of such types of flows [24,25].
%\cite{nia00}, \cite{qg00}.
In the presence of convection, the height-integrated angular momentum
and energy equations can be written as follows,
\begin{equation}
J_{adv}+J_{visc}+J_{conv}=0,
\end{equation}
\begin{equation}
\rho R T{ds\over dR}+{1\over R^2}{d\over dR}(R^2 F_{conv})=
\Omega(J_{visc}+J_{conv}),
\end{equation}
where $J_{adv}$, $J_{visc}$ and $J_{conv}$ are the 
advective, viscous and convective angular momentum fluxes respectively.
Solutions of equations (5) and (6), together with the equation of motion in
the radial direction, crucially depend on the direction of 
convective angular momentum transport. 
There are two extreme possibilities. The first one is that convection behaves
like normal viscosity, 
\begin{equation}
J_{conv}=-\nu_{conv}\rho R^3(d\Omega/dR).
\end{equation}
An alternative possibility is that the convective flux is directed down
the specific angular momentum gradient, 
\begin{equation}
J_{conv}=-\nu_{conv}\rho R[d(\Omega R^2)/dR].
\end{equation}
For $\Omega\propto R^{-3/2}$, convective fluxes (7) and (8)
correspond to outward and inward transport of 
angular momentum respectively. 
Numerical simulations have demonstrated that $J_{conv}<0$
in CDAFs.

Choosing the convective flux (8) one can construct a {\it nonaccreting}
solution with $v_R=0$. We refer to it as a ``convective envelope''.
For this solution, from equation (5) one obtains $J_{adv}=0$ and
$J_{conv}=-J_{visc}$. The latter means that in the convective envelopes
the net angular momentum flux vanishes:
the inward convective flux
is exactly balanced by the outward viscous flux.
Because of absence of advection (due to $v_R=0$) 
and zero local dissipation rate (due to zero net angular momentum flux),
the energy equation (6) is satisfied trivially
with $F_{conv}(R)\propto R^{-2}$. 
Finally, the convective envelope solution has the following scalings:
\[
 \rho(R)\propto R^{-1/2},
\]
\[
 v_R(R)=0,
\]
\begin{equation}
 \Omega(R)\propto R^{-3/2},
\end{equation}
\[
 c_s(R)\propto R^{-1/2},
\]
\[
 H(R)\propto R,
\]

The outward energy flux $F_{conv}$ corresponds to luminosity
$L_{conv}=4\pi RHF_{conv}$, which is independent of radius.
Source of this energy flux is formally located at $R=0$, but the nature of the
source is not specified in the self-similar solution. In more realistic
CDAFs, the mass accretion rate $\dot{M}$ is small, but it is not exactly zero.
The fraction of the binding
energy of the accreted mass released in the innermost region
is the source of the energy required to support convection.
In the case of CDAFs the convective luminosity can be expressed 
in the following form,
$L_{conv}=\varepsilon\dot{M}c^2$,
where the parameter $\varepsilon$ has been estimated in numerical simulations,
$\varepsilon\simeq 0.01$. A non-zero $\dot{M}$ leads to a finite 
$v_R=\dot{M}/(4\pi RH\rho ) \propto R^{-3/2}$. The latter scaling and
the scalings of other quantities in (9) agree quite well with the numerical
results (see Figure~3).

\section*{RECENT MHD SIMULATIONS}

There is good reason to believe that ``viscosity'' in differentially
rotating accretion flows is produced by magnetic stress generated by
the magneto-rotational instability (MRI) \cite{bh91}.
MHD simulations of accretion flows under conditions, in which
CDAFs could be formed, have been recently performed
%(\cite{haw00}, \cite{mhm00}, \cite{mmm00} 
([27--32] and Matsumoto et al. in this Proceedings). 
A general conclusion of these studies is that MRI,
and probably other instabilities of magnetized medium, leads to
development of turbulence in accretion flows. The turbulence re-distributes
angular momentum in the flows 
with an effective $\alpha_{eff}\sim 0.01 - 0.1$.
The simulations show a tendency for density profiles in accretion flows
to be flatter, than those in ADAFs.

In general, these results of MHD simulations do not contradict
the results of hydrodynamical simulations of convective flows,
which were discussed in previous sections.
The MHD simulations have demonstrated the range of $\alpha_{eff}$,
in which hydrodynamical models are convectively unstable.
The flattened density profiles in MHD models are in good agreement
with those found in convective hydrodynamical models.
However, a direct comparison of results of both approaches
meets difficulties.
The main problem is a low spatial resolution, and as a consequence,
a small radial range of flow patterns studied in 
the discussed MHD simulations. In this case of small radial ranges 
the effects of boundaries become significant enough to influence
the flow structure. Thus, more powerful computers and 
more developed numerical codes are required to obtain a comparable
spatial resolution to that used in hydrodynamical simulations.

Another problem which can introduce difficulties for the comparison
is that some MHD simulations have uncontrolled losses
of energy due to numerical reconnection of magnetic lines. 
These energy losses can artificially suppress convection in the 
considered flows and change the flow structures, independently of the
used spatial resolution.
The problem could be solved by using an ``artificial''
resistivity in MHD schemes (see \cite{sp00} for detailed discussion).

\section*{CONCLUSIONS}

\begin{itemize}

\item Radiatively inefficient hydrodynamical
black hole accretion flows with small
viscosity ($\alpha\la 0.1$) are always dominated by convection.
Thus for low viscosity flows the concept of ADAF is unphysical.

\item At present, CDAFs provide the best theoretical model for
understanding of radiatively inefficient, low viscosity accretion flows.

\item CDAFs are hot, their thermal energy is close to the
virial energy. They have relatively reduced mass accretion rate.
The density distribution in CDAFs is
flattened, comparing to the one in ADAFs and Bondi-like flows
($\rho\propto R^{-1/2}$ in the former case vs. 
$\rho\propto R^{-3/2}$ in the latter case).
Convection transports outward a significant amount (up to $\sim 1\%$)
of dissipated binding energy of accretion flows.
No powerful outflows are present in CDAFs.

\item CDAF is a very simple model which is now undergoing a rapid development.
A more detailed account of plasma, magnetic, and radiative processes
is likely to change this model.

\end{itemize}

\section*{ACKNOWLEDGMENTS}
 
The authors thank NORDITA and
I.V.I. thanks Harvard-Smithsonian Center for Astrophysics for hospitality
while part of this work was done. This work was supported
by NSF grant PHY 9507695, the Royal Swedish Academy of Sciences, and
RFBR grant 00-02-16135.

\end{document}